\documentclass[amsfonts,amssymb,twocolumn,notitlepage,showpacs,nofootinbib]{revtex4-1}

\pdfoutput=1

\usepackage{amsmath}
\usepackage{amsfonts}
\usepackage{amssymb}
\usepackage{mathrsfs}
\usepackage{amsthm}
\usepackage{wasysym}
\usepackage[usenames]{color}
\usepackage[usenames]{xcolor}
\usepackage[unicode, colorlinks=true, linkcolor=blue, citecolor=gray, pdfusetitle]{hyperref}
\usepackage[all]{hypcap}
\usepackage{graphicx}
\usepackage{xspace}
\usepackage{verbatim}
\usepackage{enumerate}
\usepackage[normalem]{ulem}

\usepackage[T1]{fontenc}
\usepackage[utf8]{inputenc}

\usepackage{microtype}

\graphicspath{{figs/}}

\newcommand{\figDeltaVU}{%
\begin{figure}[t]
  \centering
  \includegraphics[width=\linewidth]{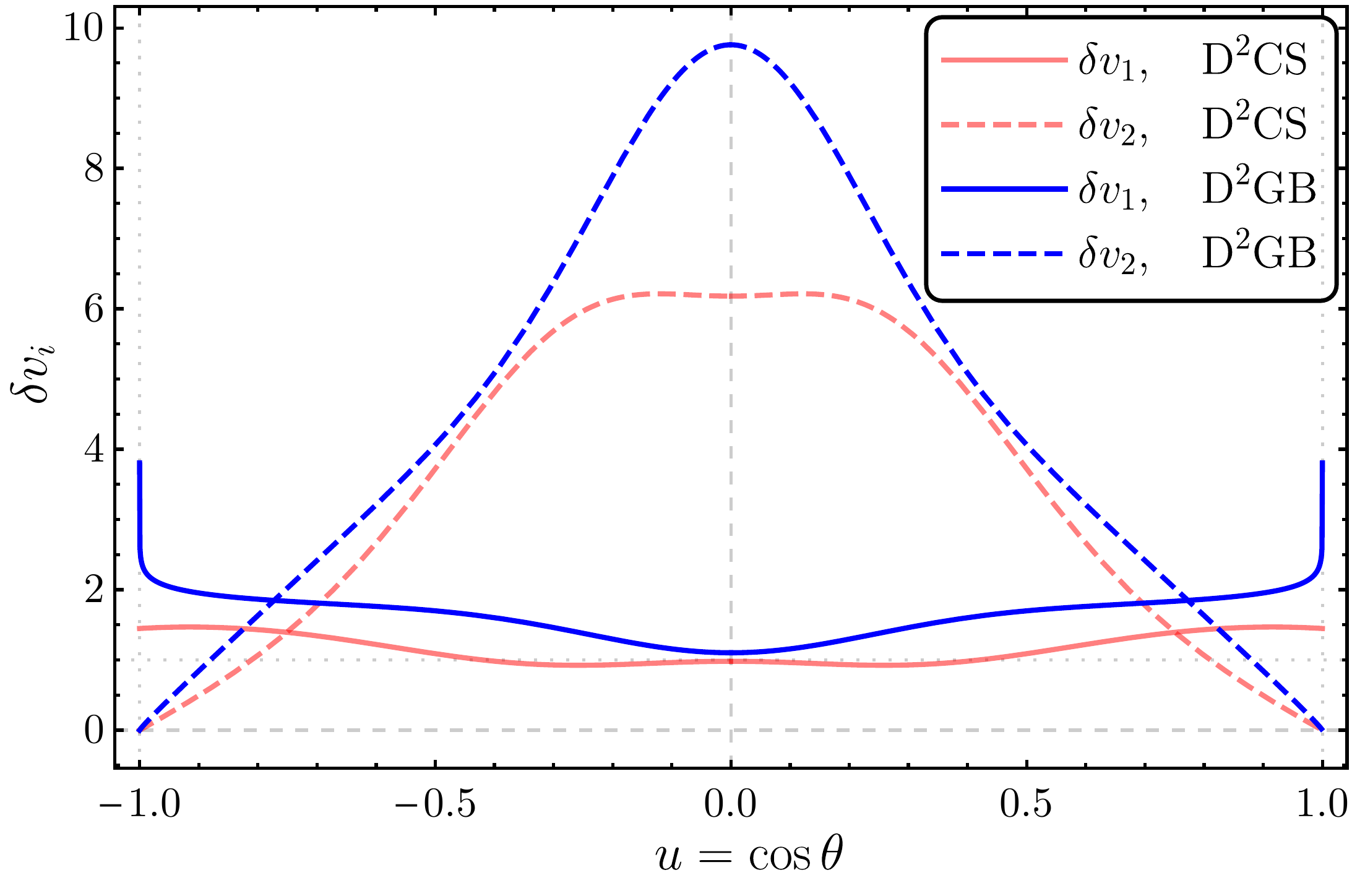}
  \caption{%
    The metric deformation functions $\delta v_1$ (solid) and
    $\delta v_2$ (dashed) as functions of $u$, for both dCS-deformed
    (red) and EdGB-deformed (blue) NHEK.  Note that in \DGB,
    $\delta v_1$ blows up at the two poles $u=\pm 1$.%
  }
  \label{fig:deltaVU}
\end{figure}
}

\newcommand{\nn}{\nonumber}

\newcommand{\pd}{\partial}

\newcommand{\txt}[1]{{\textrm{\tiny{#1}}}}

\newcommand{\mpl}{m_\txt{pl}}

\newcommand{\pont}{{}^{*}\!RR}
\newcommand{\euler}{{}^{*}\!R{}^{*}\!R}
\newcommand{\hatpont}{{}^{*}\!\hat{R}\hat{R}}
\newcommand{\hateuler}{{}^{*}\!\hat{R}{}^{*}\!\hat{R}}

\newcommand{\SLTR}{\ensuremath{SL(2,\mathbb{R})}}
\newcommand{\SLTRU}{\ensuremath{SL(2,\mathbb{R})\times U(1)}}
\newcommand{\sltr}{\ensuremath{\mathfrak{sl}(2,\mathbb{R})}}
\newcommand{\sltru}{\ensuremath{\mathfrak{sl}(2,\mathbb{R})\times \mathfrak{u} (1)}}

\newcommand{\DCS}{\ensuremath{\mathrm{D^2CS}}}
\newcommand{\DGB}{\ensuremath{\mathrm{D^2GB}}}

\newcommand{\CS}{\mathrm{CS}}
\newcommand{\GB}{\mathrm{GB}}
\newcommand{\ldcs}{\ensuremath{\ell_{\mathrm{CS}}}}
\newcommand{\lgb}{\ensuremath{\ell_{\mathrm{GB}}}}

\newcommand{\poinc}{Poincar\'{e}}

\begin{document}

\title{%
  Deformation of extremal black holes from stringy interactions}
\author{Baoyi Chen}
\email{baoyi@tapir.caltech.edu}

\author{Leo C.~Stein}
\email{leostein@tapir.caltech.edu}

\affiliation{%
  TAPIR, Walter Burke Institute for Theoretical Physics,
  California Institute of Technology, Pasadena, California 91125, USA}

\hypersetup{pdfauthor={Chen and Stein}}

\date{\today}

\begin{abstract}
  Black holes are a powerful setting for studying general relativity
  and theories beyond GR.
  However, analytical solutions for rotating black holes in beyond-GR
  theories are difficult to find because of the complexity of such
  theories.
  In this paper, we solve for the deformation to the near-horizon
  extremal Kerr metric due to two example string-inspired beyond-GR
  theories: Einstein-dilaton-Gauss-Bonnet, and dynamical Chern-Simons
  theory.
  We accomplish this by making use of the enhanced symmetry group of
  NHEK and the weak-coupling limit of EdGB and dCS.
  We find that the EdGB metric deformation has a curvature
  singularity, while the dCS metric is regular.
  From these solutions we compute orbital frequencies, horizon areas,
  and entropies.
  This sets the stage for analytically understanding the microscopic
  origin of black hole entropy in beyond-GR theories.
\end{abstract}

\maketitle

\section{Introduction}
\label{sec:introduction}

General relativity (GR), despite its huge success in describing
gravity on large scales~\cite{Will:2014kxa}, must be corrected at high
energies to reconcile with quantum mechanics.  Black holes (BHs) may
hold a key to developing a quantum theory of gravity: quantum effects
can become important when gravity is strong, such as close to
singularities.  Quantum effects can also become important at the horizon
over sufficiently long times, e.g.~as Hawking
radiation~\cite{Hawking:1974sw} shrinks a BH, generating arbitrarily
large curvatures at the horizon, close to evaporation.

In order to go beyond GR, a huge class of alternative theories of
gravity have been proposed and studied.  Analytical black hole
solutions can be sensitive to corrections to GR, but they are rare in
beyond-GR theories.  In the slow-rotation limit, BH
solutions~\cite{Yunes:2009hc, Yagi:2012ya} have been found for
dynamical Chern-Simons theory~\cite{Alexander:2009tp}.  But for many
other theories or when it comes to generic spin, it is difficult to
find analytic rotating solutions.  In this paper we find BH
solutions in the near-horizon extremal limit for beyond-GR theories.
In particular, we make use of two theories of gravity as examples,
taking the weak-coupling limit, and find the corresponding
deformations to
near-horizon extremal Kerr (NHEK).  The two theories, both
inspired by string theory, are Einstein-dilaton-Gauss-Bonnet
(EdGB)~\cite{Metsaev:1987zx, Maeda:2009uy} and dynamical Chern-Simons
theory (dCS)~\cite{Alexander:2009tp} respectively.  They both contain
a dynamical scalar field that couples to curvature, correcting GR
with a (different) quadratic curvature term.

After taking the weak coupling limit of a beyond-GR theory, finding the
vacuum rotating solutions can be naturally formulated as finding the
metric deformations to solutions in Einstein gravity,
i.e.~deformations to Kerr black holes (alternatively, one may
expand Kerr around the $a=0$ Schwarzschild limit, and solve for
deformations around the expanded spacetime, as in~\cite{Kanti:1995vq,
  Pani:2009wy, Pani:2011gy}).
Therefore linear metric perturbation theory is a natural
tool to address the problem.  However, the perturbation equations are
hard to solve unless we can use separation of variables.  In the Kerr
spacetime, metric perturbations do not separate, but curvature
perturbations do.  The most common approach is to use the
Newman-Penrose formalism~\cite{Newman:1961qr} and solve the wave
equations for Weyl scalars $\Psi_4$ or $\Psi_0$.  This method was
developed by Teukolsky~\cite{Teukolsky:1972my, Teukolsky:1973ha}, and
the partial differential equation to solve is known as the Teukolsky
equation.  The cost of curvature perturbations, however, is a very
complicated metric-reconstruction procedure (see e.g.~discussion
in~\cite{Teukolsky:2014vca}), which only works for certain source
terms, in certain gauges, and does not recover all pieces of the
metric.  The main difficulty in the separation
of the metric perturbation equations is insufficient symmetry in the
Kerr spacetime.  In the near-horizon extremal scaling limit of Kerr,
additional symmetries arise, and we can separate variables, as the
authors showed in~\cite{Chen:2017ofv}.  Therefore in NHEK, analytical
deformed solutions can be found by using linear metric perturbation
theory.

The NHEK spacetime is interesting to study for several other reasons.
For instance, it has been shown that the horizon instability of
extremal black holes~\cite{Aretakis:2012ei} can be viewed as a
critical phenomenon~\cite{Gralla:2017lto}.  Moreover, it was shown
that near-horizon quantum states can be identified with a
two-dimensional conformal field theory (CFT), via the proposed
Kerr/CFT correspondence~\cite{Guica:2008mu}.

In this paper, we focus on finding metric deformations of NHEK due to
dCS and EdGB interactions in the decoupling limit.  Let us emphasize, though, that
this formalism is not limited to these two theories, but can be
applied to finding deformed NHEK solutions in many beyond-GR theories
in the decoupling limit.  With the metric solutions, we compute
physical properties including geodesic motion of particles and their
orbital frequencies, horizon areas, and entropies.  We also prove that
the EdGB extremal BH is indeed singular in the decoupling limit,
confirming the conjecture of~\cite{Kleihaus:2015aje}.  One of the most
important results is the calculation of the macroscopic extremal
black hole entropies in beyond-GR theories.  Although we only consider
the near-horizon limit, the entropy results agree with extremal BH
solutions (i.e.~without zooming into the near-horizon region).
In the NHEK spacetime, the entropy can be computed by counting
the microscopic states of a two-dimensional chiral
CFT~\cite{Guica:2008mu} via the Cardy formula, which leads to the
Kerr/CFT conjecture.  We also expect a dual CFT description
of the extremal black hole entropy for beyond-GR theories in the
decoupling limit.  We will not address this issue here, but our work
lays the ground for studying the microscopic states of deformed
extremal black holes.  This may provide insight into quantum theories
beyond Einstein gravity.

We organize the paper as follows.
In Sec.~\ref{sec:edgb-and-dcs} we review EdGB and dCS gravity, and
introduce the decoupling limit to the two theories.
In Sec.~\ref{sec:kerr-nhek-limit}, we review the near-horizon extremal
geometry, the symmetry-adapted bases, and set up the metric
perturbations in near-horizon extremal Kerr spacetime as induced by
the two stringy interactions.
In Sec.~\ref{sec:solve-met-def} we solve for the dynamical scalar fields,
construct the source term to the linearized Einstein field equation,
and finally solve the metric perturbations in the ``attractor'' gauge.
In Sec.~\ref{sec:properties-solutions} we derive the timelike
geodesic equations for the deformed spacetimes, and calculate the
corrections to horizon areas and black hole entropies due to the two
stringy interactions.
We conclude and discuss future work in
Sec.~\ref{sec:concl-future-work}.

\section{Einstein-dilaton-Gauss-Bonnet and dynamical
  Chern-Simons gravity}
\label{sec:edgb-and-dcs}

\subsection{Action}
\label{sec:the-action}

We work in units where $c=1=\hbar$, and choose the metric signature
$(-,{}+,{}+,{}+)$.
The theories which we are considering, namely dynamical Chern-Simons
gravity and Einstein-dilaton-Gauss-Bonnet, can be motivated from both
low-energy effective field theory (EFT) and high-energy fundamental theory.
DCS can arise from gravitational anomaly cancellation in chiral
theories~\cite{Delbourgo:1972xb, Eguchi:1976db, AlvarezGaume:1983ig},
including Green-Schwarz cancellation in string
theory~\cite{Green:1984sg}.  The low-energy compactified theory was
explicitly presented in~\cite{Campbell:1990fu} (and see references
therein).  EdGB, meanwhile, can be derived by expanding the low energy
string action to two loops to find the dilaton-curvature
interaction~\cite{Metsaev:1987zx, Maeda:2009uy}.

The actions of dCS and EdGB both include the Einstein-Hilbert term and
a scalar field that non-minimally couples to curvature.  The
Einstein-Hilbert action leads to standard GR.  In dCS the scalar field
is an axion, while in EdGB it is a dilaton.  In our discussions, there
is no need to distinguish between the two scalar fields.  We treat
them equally as the scalar field $\vartheta$.  For both theories, we
then take as our action
\begin{equation}
  I = \int d^4x\sqrt{-g}\left[\mathscr{L}_{\mathrm{EH}}
    +\mathscr{L}_{\vartheta}+\mathscr{L}_{\mathrm{int}}\right],
  \label{eq:the-action}
\end{equation}
with
\begin{align}
  \mathscr{L}_{\mathrm{EH}} &= \frac{1}{2}\mpl^2R\,,
  & \mathscr{L}_{\vartheta}
  &= -\frac{1}{2}(\partial^a\vartheta)(\partial_a\vartheta)\,,
\end{align}
and non-minimal scalar-curvature interaction terms for dCS and EdGB
respectively~\cite{Alexander:2009tp, Metsaev:1987zx, Maeda:2009uy}
\begin{align}
  \mathscr{L}^{\CS}_{\mathrm{int}}
  &= -\frac{\mpl}{8}\ldcs^2\vartheta\,\pont\,, &
     \mathscr{L}^{\GB}_{\mathrm{int}}
  &= -\frac{\mpl}{8}\lgb^2\vartheta\,\euler \,.
\end{align}
Here $R$ is the Ricci scalar of the metric $g_{ab}$, and $g$ is the
metric determinant.  The reduced Planck mass is defined through
$\mpl\equiv(8\pi G)^{-1/2}$. The scalar field $\vartheta$ has been
canonically normalized such that $[\vartheta]=[M]$.  In the
interaction terms, we define two coupling constant $\ldcs$ and $\lgb$
for dCS and EdGB respectively.  The two variables are dimensionful,
specifically $[\ldcs]=[\lgb]=[M]^{-1}$. That is, each of them gives
the length scale of the corresponding theory, which in principle can
be constrained observationally.  In dCS, we encounter the
Pontryagin-Chern density
\begin{equation}
  \pont = {}^{*}\!R^{abcd}R_{abcd}
  \,,
\end{equation}
while in EdGB we see minus the Euler (or Gauss-Bonnet) density
\begin{equation}
  \euler = {}^{*}\!R^{*}_{abcd}{}R^{abcd}=-R^2+4R_{ab}R^{ab}-R_{abcd}R^{abcd}
  \,.
\end{equation}
Here we have used the single- and double-dualized Riemann tensors,
\begin{align}
  \label{eq:R-duals}
  {}^{*}\!R_{abcd} &\equiv
  \frac{1}{2} \epsilon_{ab}{}^{ef} R_{efcd}
  \,, &
  {}^{*}\!R^{*}_{abcd} &\equiv
  \frac{1}{2} {}^{*}\!R_{abef} \epsilon^{ef}{}_{cd}
  \,,
\end{align}
where we dualize with the completely antisymmetric Levi-Civita tensor
$\epsilon^{abcd}$.

\subsection{Equation of motion}
\label{sec:eq-of-motion}

Variation of the action in Eq.~\eqref{eq:the-action} with respect to
the scalar field $\vartheta$ leads to the scalar equation of motion
for dCS and EdGB respectively,
\begin{align}
  \square \vartheta & = \frac{\mpl}{8}\Bigg\{    \begin{array}{l}
      \ldcs^2\,\pont\,, \\
      \lgb^2\,\euler\,, \\
    \end{array}  &
     \begin{array}{cc}
       \text{dCS}\\
       \text{EdGB}
     \end{array}
\end{align}
where $\square=\nabla^a\nabla_a$ and $\nabla_a$ is the covariant
derivative compatible with the metric. Variation of the action in
Eq.~\eqref{eq:the-action} with respect to $g^{ab}$ leads to the metric
equation of motion,
\begin{align}
  \mpl^2G_{ab}
  &= T_{ab}[\vartheta,\vartheta]
  - \mpl
  \Bigg\{
  \begin{array}{cc}
    \ldcs^2 C_{ab}[\vartheta]\,, \\
    \lgb^2 H_{ab}[\vartheta]\,,
  \end{array}
  &
    \begin{array}{cc}
      \text{dCS}\\
      \text{EdGB}
    \end{array}
\end{align}
Here $T_{ab}[\vartheta,\vartheta]$ is the canonical
stress-energy tensor for the scalar field $\vartheta$,
\begin{equation}
  T_{ab}[\vartheta,\vartheta]=\nabla_a \vartheta \nabla_b\vartheta - \frac{1}{2}g_{ab}\nabla^c\vartheta\nabla_c\vartheta
  \,.
\end{equation}
We also define the $C$-tensor for dCS, 
\begin{equation}
  C_{ab}[\vartheta]  =
  \nabla^c \nabla^d\left[{{}^{*}\!R}_{d(ab)c} \vartheta \right]
  \,,
\end{equation}
and introduce the $H$-tensor for EdGB via
\begin{align}
  H_{ab}[\vartheta] =
    \nabla^c \nabla^d\left[{}^{*}\!R^*_{dabc} \vartheta \right]
    \,,
\end{align}
where parentheses around $n$ indices means symmetrizing with a factor
of $1/n!$.

\subsection{Decoupling limit}
\label{sec:the-decouple-limit}

We now introduce two distinct theories as the decoupling limit of dCS
and EdGB respectively, namely \textit{Decoupled dynamical
  Chern-Simons} (\DCS) and \textit{Decoupled dynamical Gauss-Bonnet}
(\DGB)~\cite{Yagi:2015oca}. We will briefly review the formalism of
taking the decoupling limit in dCS (see~\cite{Stein:2014xba} for
detailed discussions). The extension of this formalism to EdGB is
straightforward.

We assume the corrections to GR due to the interaction terms are
small, so that in the limit $\ell \to 0$, we recover standard GR.
This allows us to perform a perturbative expansion of
all the fields in terms of powers of $\ldcs$.  To make the
perturbation theory simpler, we introduce a formal dimensionless
order-counting parameter $\varepsilon$.  We then consider a
one-parameter family of theories defined by the action
$I_{\varepsilon}$, where in $I_{\varepsilon}$, we have multiplied
$\mathscr{L}_{\mathrm{int}}$ by $\varepsilon$.  This parameter can be
set to 1 later.

Now we expand all fields and equations of motion in a series
expansion in powers of $\varepsilon$.  Specifically, we take
$\vartheta = \vartheta^{(0)} + \varepsilon \vartheta^{(1)} +
\mathcal{O}(\varepsilon^{2})$, and similarly
$g_{ab} = g_{ab}^{(0)} + \varepsilon h_{ab}^{(1)} + \varepsilon^{2}
h_{ab}^{(2)} + \mathcal{O}(\varepsilon^{3})$.

In order to
recover GR in the limit $\varepsilon\rightarrow 0$, at order
$\varepsilon^0$, we have $\vartheta^{(0)}=0$.  At order
$\varepsilon^1$, $h_{ab}^{(1)}$ has vanishing source term and thus can
be set to zero as well.  It is then easy to show that the EOM for
the leading order scalar field $\vartheta^{(1)}$ is at
$\varepsilon^1$, given by
\begin{equation}
  \square^{(0)} \vartheta^{(1)} = \frac{\mpl}{8}\ldcs^2\,[\pont]^{(0)}\,,
\end{equation}
and the leading order metric deformation enters at
$\varepsilon^2$, which satisfies
\begin{equation}
  \mpl^2G^{(1)}_{ab}[h^{(2)}] + \mpl \ldcs^2 C_{ab}[\vartheta^{(1)}]
  =
  T_{ab}[\vartheta^{(1)},\vartheta^{(1)}]
  \,.
\end{equation}
Here $G^{(1)}_{ab}[h^{(2)}]$ is the linearized Einstein operator
acting on the metric deformation $h^{(2)}_{cd}$.

We now redefine our field variables in powers of $\ldcs$, but to do so
we need another length scale against which to compare.  This
additional length scale is given by the typical curvature radius of
the background solution, e.g.~$L \sim |R_{abcd}|^{-1/2}$.  For a
black hole solution, this length scale will be $L\equiv GM$.
We can then also pull out the scaling with powers of $L$ from spatial
derivatives and curvature tensors, by defining $\hat{\nabla}=L\nabla$
and $\hat{R}_{abcd}=L^{2}R_{abcd}$.
We define $\hat{h}_{ab}$ and $\hat{\vartheta}$ via
\begin{align}
  \vartheta^{(1)} &= \mpl
                    \left(
                    \frac{\ldcs}{GM}
                    \right)^2\hat{\vartheta}\,,
  &
  h^{(2)}_{ab}
  &=
    \left(
    \frac{\ldcs}{GM}
    \right)^{4}\hat{h}_{ab}
  \,.
\end{align}
Now our hatted variables satisfy the dimensionless field equations
\begin{align}
  \hat{\square}^{(0)} \hat{\vartheta} &= \frac{1}{8} [\hatpont]^{(0)} \,, &
  G^{(1)}_{ab}[\hat{h}] = S_{ab}
  \,,
  \label{eq:linearized-Einstein}
\end{align}
with the source term
$ S_{ab}=
T_{ab}[\hat{\vartheta},\hat{\vartheta}] -
C_{ab}[\hat{\vartheta}] $.

The equations of motion in
the decoupling limit of EdGB, i.e.~\DGB, are almost the same as
Eq.~\eqref{eq:linearized-Einstein}.  The only difference is that,
for EdGB, we substitute $\hateuler$ for $\hatpont$, and the $C$-tensor in
the source term should be replaced by the $H$-tensor.

\section{NHEK and separable metric perturbations}
\label{sec:kerr-nhek-limit}

The metric of a generic near-horizon extremal geometry (NHEG) that makes
\SLTRU{} symmetry manifest takes the form~\cite{Astefanesei:2006dd}
\begin{align}
  \label{eq:NHEG-met}
  \text{d}s^2 ={} &  (GM)^2
                  \bigg[v_1(\theta)\left( -r^2\,\text{d}t^2 +                 \frac{\text{d}r^2}{r^2}  + \beta^2\text{d}\theta^2 \right) \\ \nn
                &  + \beta^2v_2(\theta)(\text{d}\phi - \alpha r\,\text{d}t)^2 \bigg]
                  \,,
\end{align}
where $v_1$ and $v_2$ are positive functions of the polar angle $\theta$,
and $\alpha$ and $\beta$ are constants.  The spacetime has four Killing
vector fields.  In these \poinc{} coordinates, they are given by
\begin{align}
  H_0&=t\partial_t -r\partial_r\,, \\ \nn
  H_+&=\partial_t\,, \\ \nn
  H_-&=(t^2+\frac{1}{r^2})\partial_t- 2tr\partial_r +\frac{2\alpha}{r}\partial_\phi\,, \\ \nn
  Q_0&=\partial_\phi \,.
\end{align}
The four generators form a representation of the Lie algebra
$\mathfrak{g} \equiv \sltru $,
\begin{align}
  \label{eq:Lie-algebra-poincare}
  &[H_0 \,, H_\pm\,] = \mp H_\pm \,, \\ \nn
  &[H_+ , \,H_-] = 2\,H_0 \,,      \\ \nn
  &[H_s \,\,, \,Q_0\,] = 0 \,. \qquad (s=0,\pm)
\end{align}
A crucial algebra element we will need is the Casimir element of
\sltr.  The Casimir $\Omega$ acts on a tensor $\bf t$ via
\begin{equation}
  \Omega \cdot {\bf t} =
  \left[\mathcal{L}_{H_0}(\mathcal{L}_{H_0}-\mathrm{id})
    - \mathcal{L}_{H_-}\mathcal{L}_{H_+} \right]{\bf t}
  \,,
\end{equation}
where $\mathcal{L}_{X}$ is the Lie derivative along the
vector field $X$.

The generic metric in Eq.~\eqref{eq:NHEG-met} has an Einstein gravity
solution, which is found with
\begin{align}
  v_1(u) &= 1+u^2\,, &
    \alpha &=-1\,, &\\ \nn
  v_2(u) &=\frac{4(1-u^2)}{1+u^2}\,, &
    \beta  &=+1\,, &
\end{align}
where we have defined a new coordinate $u=\cos\theta$. This spacetime
is called near-horizon extremal Kerr, which was first obtained by
taking the near-horizon limit of extremal Kerr black
holes~\cite{Bardeen:1999px}.

The enhanced symmetry due to the near-horizon extremal limit enables
us to separate variables in the linearized Einstein equation (LEE) in NHEK
spacetime~\cite{Chen:2017ofv}.  This is achieved by expanding the
metric perturbations in terms of some basis functions adapted to that
symmetry.  For the non-compact group \SLTR, one can construct a
\textit{highest-weight module}, which is a unitary irreducible
representation of the group.  In NHEK, that is, we simultaneously
diagonalize $\{\mathcal{L}_{Q_0},\Omega,\mathcal{L}_{H_0}\}$ and label
the eigenfunctions $\xi$ by $m,h,k$ respectively.  Here $m$ labels the
azimuthal direction, $h$ labels the representation (``weight''), and
$k$ labels ``descendants'' within the same representation.
We impose the highest-weight condition
$\mathcal{L}_{H_+}\xi=0$, and solve for the basis functions. Expanding
the metric perturbations in terms of these bases leads to separation of
variables for the LEE in NHEK spacetime.  As a result, the system of
partial differential equations in the LEE automatically turns into one
of ordinary differential equations.

If the LEE system has a source term, and that source term is a linear
combination of a finite number of representations, then
the metric perturbations can also be expanded as a sum of those same
representations.
As we will see, for both EdGB and dCS gravity in the
decoupling limit, the source term on the RHS of
Eq.~\eqref{eq:linearized-Einstein} will have the same \SLTRU~symmetry
as the background spacetime.  This enables us to solve for the linear
metric deformations analytically.

\section{Solving for the metric deformations}
\label{sec:solve-met-def}

In this section we find solutions of the leading order
scalar fields, construct the source terms on the RHS of
Eq.~\eqref{eq:linearized-Einstein} for \DCS{} and \DGB{} respectively,
and finally solve for the metric deformations.

\subsection{Solutions for scalars and construction of source}
\label{sec:solut-scal-constr}

In a Ricci-flat spacetime (like Kerr), the $I$ curvature
invariant~\cite{Stephani:2003ika} agrees with $I=\frac{1}{16}(-\euler
+ i \pont)$.  In NHEK, this takes the value $\hat{I}=3/(1-iu)^{6}$.
The imaginary and (minus) real parts of $\hat{I}$ thus give compact
ways of expressing the source terms for the scalar equations of motion
of respectively \DCS{} and \DGB{}.

In \DCS{}, the leading order scalar equation of motion admits an
axion solution which is regular everywhere. This scalar field
is given by
\begin{equation}
  \hat{\vartheta}^{(1)}=
  \frac{1}{4}\left[\frac{u \left(u^4+2 u^2-7\right)}
    {\left(u^2+1\right)^3}+ 2\arctan u \right] +
  \mathrm{const}
  \,.
\end{equation}
This also agrees with the solution presented in~\cite{McNees:2015srl}.
Because the theory is shift-symmetric, we are free to set the constant
term to zero.  We then construct the source
$S_{ab}[\hat{\vartheta}^{(1)},\hat{\vartheta}^{(1)}]$ in
Eq.~\eqref{eq:linearized-Einstein} for \DCS.

In \DGB{}, we find the leading order scalar solution is
\begin{align}
  \hat{\vartheta}^{(1)}
  ={}& d_2+\frac{\log \left(u^2+1\right)}{4}
    -\frac{u^4+4 u^2-1}{2\left(u^2+1\right)^3} + \\ \nn
  &+\left(-\frac{d_1}{2} -\frac{1}{4}\right)\log (1-u) \,+  \\ \nn
  &+\left(\frac{d_1}{2}-\frac{1}{4}\right) \log (1+u)
    \,,
\end{align}
where $d_1$ and $d_2$ are constants.  Unlike the \DCS{} case,
it is not possible to remove both logarithmic divergences at $u=\pm1$
by choosing specific values of $d_1$ and $d_2$.  It is possible to
cancel the divergence at one pole or the other, but not both.
We set $d_1=0$ so that the
scalar field retains the reflection symmetry, $u\to -u$, of the
background spacetime.  Again by shift symmetry, we are free to set the
additive constant $d_2=0$, and then construct the source term $S_{ab}$
accordingly.  The source $S_{ab}$ remains irregular at the two poles
$u=\pm1$.

Let us remark on an important common feature of the two source
terms.  For either theory,
\begin{equation}
  \mathcal{L}_{X}S_{ab}=0
  \,,
\end{equation}
where $X\in\{H_0,H_\pm,Q_0\}$.  That is, if we decompose the source
term using the symmetry-adapted scalar, vector and tensor bases, the
source term only contains the $m=h=k=0$ component.  Therefore on the
LHS of the LEE, the metric perturbations only have stationary
axisymmetric basis components, either for \DCS\, or
\,\DGB.  These components live in both the highest-weight and
lowest-weight representations of NHEK's isometry group.

\subsection{dCS-deformed NHEK}

We now seek the solutions to the linearized metric perturbation
equations of NHEK sourced by the two stringy interactions.
Expansions of the metric perturbations into the basis functions turn the systems of
partial differential equations in LEE into ten coupled ordinary
differential equations (ODEs) in $u$, which we solve in this subsection.

So far we haven't chosen any gauge condition.  Since the linear metric
perturbations have the same \SLTRU{} symmetry as the background NHEK
spacetime, we can fix the gauge by requiring an ``attractor form''~\cite{Astefanesei:2006dd} of
the deformed solutions as in Eq.~\eqref{eq:NHEG-met}.  That is, we
only consider the following shifts in the metric parameters.  Recalling
that the metric is corrected at order $\varepsilon^2$,  we
have 
\begin{align}
  v_1(u)&\rightarrow v_1(u)+ \varepsilon^2\delta v_1(u)\,, & \alpha\rightarrow \alpha + \varepsilon^2\delta\alpha\,, &\\ \nn
  v_2(u)&\rightarrow v_2(u)+ \varepsilon^2\delta v_2(u)\,, & \beta\rightarrow \beta + \varepsilon^2\delta\beta\,. &
\end{align}
We call this gauge choice the \textit{attractor gauge}.
This ansatz is, by construction, in the $m=h=k=0$ representation of
NHEK's isometry group.  Therefore it always makes the \SLTRU{} symmetry manifest.

\figDeltaVU{}

For \DCS, the linear metric deformations are found to be the following
complicated expressions, which we also plot in Fig.~\ref{fig:deltaVU}:
\begin{align}
  &\delta v_1(u)= f_1(u)+  \frac{1}{53760 \left(u^2+1\right)^5}\mathcal{P}^{\DCS}_{1}[u]
    \,, \\
  &\delta v_2(u) = f_2(u)  -\frac{\left(u^2-1\right)}{6720
    \left(u^2+1\right)^7}  \mathcal{P}^{\DCS}_{2}[u]
    \,,
\end{align}
where
\begin{align}
\label{eq:f1}
  &f_1(u)=\frac{1}{3} c_1 \left(-u^2+4 u-1\right)+\frac{1}{3} c_2 \left(2 u^2-5 u+2\right) \\ \nn
  & -\frac{1}{3} c_3 u \sqrt{1-u^2}
    -\frac{4}{3} \delta \beta  \left(u^2+1\right)+2 \delta \beta  u \sqrt{1-u^2} \sin ^{-1}u \\ \nn
  & +\frac{975 u \sqrt{1-u^2} \tan ^{-1}\left(\frac{\sqrt{2} u}{\sqrt{1-u^2}}\right)}{512 \sqrt{2}}-\frac{3}{16} u \tan ^{-1}u \,,
\end{align}
\begin{align}
\label{eq:f2}
  &f_2(u)  = \frac{8 c_3 u \sqrt{1-u^2}}{3 \left(u^2+1\right)^2}+\frac{4 c_1 \left(u^4+4 u^3-4 u-1\right)}{3 \left(u^2+1\right)^2} \\ \nn
  & -\frac{4 c_2 \left(2 u^4+5 u^3-5 u-2\right)}{3 \left(u^2+1\right)^2}+\frac{40 \delta \beta  \left(u^2-1\right)}{3 \left(u^2+1\right)} \\ \nn
  &  -\frac{16 \delta \beta  u \sqrt{1-u^2} \sin ^{-1}u}{\left(u^2+1\right)^2}+\frac{\delta\alpha  \left(8-8 u^2\right)}{u^2+1}
  \\ \nn
  &-\frac{975 u \sqrt{1-u^2} \tan ^{-1}\left(\frac{\sqrt{2} u}{\sqrt{1-u^2}}\right)}{64 \sqrt{2} \left(u^2+1\right)^2}-\frac{3 u \left(u^2-1\right) \tan ^{-1}u}{4 \left(u^2+1\right)^2} \,,
\end{align}
and the polynomials $\mathcal{P}^{\DCS}_{1}[u]$ and
$\mathcal{P}^{\DCS}_{2}[u]$ are given by
\begin{align}
 &\mathcal{P}^{\DCS}_{1}[u] = -58501 u^{12}  -222147 u^{10}-255058 u^8
  \\ \nn
  &\quad+11754 u^6+323735 u^4
    -149799 u^2+4416 \,, \\
 &\mathcal{P}^{\DCS}_{2}[u] = 280 u^{12}
    -52341 u^{10}-252928 u^8 \\ \nn
 & \quad-472090 u^6-536680 u^4
   +26583 u^2-18792\,.
\end{align}
Here $c_1$, $c_2$ and $c_3$ are integration constants.  It
is straightforward to see these three constants, together with
$\delta\alpha$ and $\delta\beta$, correspond to different homogeneous
solutions to the LEE.
These solutions are finite on the domain $u\in[-1,+1]$, but would have
infinite derivative at the poles $u=\pm1$ without an appropriate
choice of $\delta\beta$.  By demanding regularity at the two poles
and reflection symmetry of the deformed metric, we set
\begin{align}
  \label{eq:dCS-constants}
  \delta \beta &= -\frac{975}{1024 \sqrt{2}}
                 \,, &   c_3&= 0\,,
  & c_2 &= \frac{4 c_1}{5} \,.
\end{align}
Note that $\delta\alpha$ will shift the Killing vector $H_-$.  By
demanding that the perturbed spacetime has the same Killing vectors as
NHEK, we also set $\delta\alpha=0$.  After inserting the solutions
from \eqref{eq:dCS-constants} back into the metric, we only need to
fix $c_1$.
Collecting the terms proportional to $c_{1}$, one immediately finds that
\begin{align}
  \bigg(\text{coefficient of }c_1\bigg) \propto \frac{\partial g^{(0)}_{ab}}{\partial M}\,.
\end{align}
This means the homogeneous solution associated with $c_1$ shifts the
mass of the black hole.  Since we don't want the mass shift, we fix
$c_1=0$.  With these parameter choices, we obtain the regular
solution to the LEE sourced by the dCS interaction in the decoupling
limit. We call the newly-found spacetime \textit{dCS-deformed NHEK}.

\subsection{EdGB-deformed NHEK}

For \DGB, in the attractor gauge, the linear metric
deformations are found to be
\begin{align}
\label{eq:deltav1GB}
  &\delta v_1(u) =f_1(u)+\frac{1}{8} \left(-u^2+u-1\right) \log (1-u) \\ \nn
  &+\frac{1}{8} \left(-u^2-u-1\right) \log (1+u) \\ \nn
  &+\frac{1}{8} \left(u^2+1\right) \log \left(u^2+1\right)-\frac{3 u \sqrt{1-u^2} \tan ^{-1}\left(\frac{\sqrt{2} u}{\sqrt{1-u^2}}\right)}{256 \sqrt{2}} \\ \nn
  &  + \frac{1}{53760 \left(u^2+1\right)^5}\mathcal{P}^{\DGB}_{1}[u]
    \,,
\end{align}
\begin{align}
\label{eq:deltav2GB}
  &\delta v_2 = f_2(u)+\frac{\left(u^4-u^3+u-1\right) \log (1+u)}{2 \left(u^2+1\right)^2} \\ \nn
  &+\frac{\left(u^4+u^3-u-1\right) \log (1-u)}{2 \left(u^2+1\right)^2}+\frac{\left(1-u^2\right) \log \left(u^2+1\right)}{2 u^2+2} \\ \nn
  & +\frac{3 u \sqrt{1-u^2} \tan ^{-1}\left(\frac{\sqrt{2}
    u}{\sqrt{1-u^2}}\right)}{32 \sqrt{2} \left(u^2+1\right)^2}
    +\frac{(u^2-1)}{6720
    \left(u^2+1\right)^7}\mathcal{P}^{\DGB}_{2}[u]
    \,,
\end{align}
where the functions $f_{1}(u), f_{2}(u)$ are identical to the \DCS{}
case and given in Eqs.~\eqref{eq:f1} and \eqref{eq:f2}; and
where the polynomials $\mathcal{P}^{\DGB}_{1}[u]$ and
$\mathcal{P}^{\DGB}_{2}[u]$ are given by
\begin{align}
  &\mathcal{P}^{\DGB}_{1}[u] = -27459 u^{12}-82773 u^{10}
   -42302 u^8  \\ \nn
  &\quad +81766 u^6-18815 u^4+298479 u^2+11264 \,, \\
  &\mathcal{P}^{\DGB}_{2}[u] = 35859 u^{10}
    +152792 u^8+226230 u^6 \\ \nn
  &\quad +10160 u^4+205503 u^2-5632\,.
\end{align}
As in the \DCS{} case, the constant $\delta\beta$ can be chosen so as
to cancel a square-root behavior at the poles which would have
infinite derivative.  However, the important difference from \DCS{} is
the appearance of $\log$ terms in Eqs.~\eqref{eq:deltav1GB} and
\eqref{eq:deltav2GB}.  There are no integration constants which can
cancel these logarithmic divergences.

Still, canceling the square-root behavior and assuming reflection
symmetry in $u$, we find
\begin{align}
  \delta \beta &= -\frac{969}{1024 \sqrt{2}}
                 \,, &   c_3&= 0\,,
  & c_2 &= \frac{4 c_1}{5} \,.
\end{align}
We also fix $\delta\alpha=0$ to preserve the Killing vector
fields of NHEK, and set $c_1=0$ to avoid a mass shift.
After fixing all constants, these functions are plotted in
Fig.~\ref{fig:deltaVU}.
We call the corresponding spacetime \textit{EdGB-deformed NHEK}.
This metric deformation has a true curvature singularity at the poles,
$u=\pm 1$, which we discuss further in
Sec.~\ref{sec:concl-future-work}.

\section{Properties of solutions}
\label{sec:properties-solutions}

\subsection{Orbits}
\label{sec:orbit}
In this subsection we derive the geodesic equations for a particle in
the deformed NHEK spacetime.  Since the NHEK background and the
deformed solutions have the same isometry group, we consider the spacetime
with the general metric in Eq.~\eqref{eq:NHEG-met}.  The relativistic
Hamiltonian for geodesic motion of a particle can be defined as
\begin{align}
  H(x^a,p_b)=\frac{1}{2}g^{ab}p_ap_b\,,
\end{align}
where $p_a$ are the conjugate momenta of the particle.  By drawing
analogy to geodesic motion in Kerr spacetime, we can similarly find
three constants of motion: energy $E\equiv -p_t$, $z$ angular momentum
$L_{z}\equiv p_\phi$, and Carter's constant $\mathcal{C}$.  The Carter
constant comes from separating the radial and polar motions.
Note, however, that because our Killing vector field $\pd_{t}$ is
different from the asymptotically timelike KVF (with norm $-1$ at
infinity), our energy is different from the usual Kerr orbital
energy~\cite{Compere:2017hsi}.
Following the Hamilton-Jacobi approach~\cite{Schmidt:2002qk}, we
define the characteristic function $W$ via
\begin{align}
  W=-\frac{1}{2}\kappa \lambda -E t+\int \frac{\sqrt{R(r)}}{\beta^2 r^2}dr + \int \sqrt{\Theta(\theta)}d\theta +L_{z} \phi \,,
\end{align}
where $\lambda$ is the affine parameter and $\frac{1}{2}\kappa$ is the
value of the Hamiltonian evaluated along the world-line of the
particle.  $R(r)$ and $\Theta(\theta)$ are given by
\begin{align}
  R(r) &= \beta^4(E-\alpha L_{z} r)^2-\beta^2 \mathcal{C} r^2 \,, \\ \nn
  \Theta(\theta) &=\mathcal{C} - \frac{v_1(\theta)}{v_2(\theta)}L_{z}^2+M^2\beta^2v_1(\theta)\kappa\,.
\end{align}
Since $p_a=\frac{\partial W}{\partial x^a}$, we obtain the following
geodesic equations of motion,
\begin{align}
  \Sigma \frac{dt}{d\lambda}&=\frac{\beta^2}{r^2}(E-\alpha L_{z} r)\,, \\ \nn
  \Sigma \frac{dr}{d\lambda}&=\pm \sqrt{R(r)}\,, \\ \nn
  \Sigma \frac{d\theta}{d\lambda}&=\pm \sqrt{\Theta(\theta)}\,, \\ \nn
  \Sigma \frac{d\phi}{d\lambda}&=\frac{\alpha \beta^2}{r} (E-\alpha L_{z} r) + \frac{v_1(\theta)}{v_2(\theta)} L_{z}\,,
\end{align}
where $ \Sigma = M^2\beta^2v_1(\theta)$.  These integrals can be
directly performed after defining the ``Mino time'' $\tau$, where
$d\tau=d\lambda/\Sigma$ (this again differs from the usual Mino time in
the asymptotic region of Kerr, because our time coordinate differs).

In particular, let us consider circular equatorial motion,
i.e.~$\theta=\pi/2=\theta_0$.  For such motion we only need $E$ and
$L_{z}$ to determine the orbit.  For a time-like orbit with
four-velocity $u^a$, $g_{ab}u^au^b=-1$, we have that
\begin{align}
  \left(\frac{dr}{d\lambda}\right)^2=V(r),
\end{align}
where the effective potential $V(r)$ is given by
\begin{align}
  V(r)=\frac{(E-\alpha L_{z} r)^2}{M^4v_1^2(\theta_0)} - \frac{r^2}{M^2v_1(\theta_0)} - \frac{L_{z}^2r^2}{M^4\beta^2v_1(\theta_0)v_2(\theta_0)}\,.
\end{align}
Solving for the conditions of circular motion, we obtain
\begin{align}
  E&=0, &  L_{z}&=\pm \frac{M\beta\sqrt{v_1(\theta_0)v_2(\theta_0)}}{\sqrt{-v_1(\theta_0)+\alpha^2\beta^2v_2(\theta_0)}} \,. &
\end{align}
The corresponding circular orbits $r=r_0$ are all marginally stable,
i.e.~$V^{''}(r)|_{r=r_0}=0$. After integrating out the azimuthal
motion we also obtain that $\phi=\phi_0+\omega_\phi t$, where the
angular frequency $\omega_\phi$ is given by
\begin{align}
\label{eq:angular-freq}  \omega_\phi=\left(\alpha-\frac{v_1(\theta_0)}{\alpha\beta^2v_2(\theta_0)}\right)r_0\,.
\end{align}
The fact that all circular equatorial orbits are essentially the same,
with a different angular frequency, is due to the dilation symmetry of
the spacetime.  That is, the metric is invariant under
$r \rightarrow cr$ and $t\rightarrow t/c$ for any constant
$c\in(0,+\infty)$.  As a result, in Eq.~\eqref{eq:angular-freq}, the
radius-frequency relationship has to be compatible with the dilation
symmetry.

Plugging in the \DCS~solutions, we find the angular frequency of the
equatorial circular orbits to be
\begin{align}
  \omega^\DCS_\phi=\left[-\frac{3}{4}+\frac{25}{128}\left(\frac{\ldcs}{GM}\right)^4+\mathcal{O}\left(\varepsilon ^3\right)\right]r_0
  \,.
\end{align}
Similarly for the \DGB~solutions, the angular frequency is found to be
\begin{align}
\omega^\DGB_\phi=\left(-\frac{3}{4}+\mathcal{O}\left(\varepsilon ^3\right)\right)r_0
  \,.
\end{align}
Therefore at the leading order in the metric perturbations, EdGB-type
interactions do not lead to corrections to the angular frequency of
circular equatorial orbits in an extremal black hole, in the
near-horizon limit.

Again, because our time differs from the time coordinate in the
asymptotic region, these frequencies are not the asymptotically
observable orbital frequencies.  Such observable quantities were
computed for slowly-rotating BHs in \DCS{} in~\cite{Yagi:2012ya} and
in \DGB{} in~\cite{Ayzenberg:2014aka, Pani:2009wy, Pani:2011gy,
  Maselli:2015tta, Cunha:2016wzk}.

\subsection{Location and area of deformed horizons}
\label{sec:locat-deform-horiz}

Since NHEK is not asymptotically flat, it does not have an event
horizon.  However, because of what the near-horizon limit is designed
to do---to zoom in on the horizon region---the scaling limit of the
Kerr event horizon gives rise to the horizon of the \poinc{} patch.
This \poinc{} horizon has the same geometric properties as in Kerr,
and thus it has the same area and entropy.

We can identify the location of this Killing horizon by considering
observers whose world lines are along real linear combinations
$c_{t} \pd_{t}+c_{\phi} \pd_{\phi}$, with $c_{t}, c_{\phi}$ real
constants, such that their world lines are timelike.  At the horizon,
these world lines are forced to be null.  For any metric of the NHEG
form~\eqref{eq:NHEG-met}, the horizon is at $r=0$.  Therefore in
attractor gauge, the coordinate location of the horizon is not
deformed after including the scalar-gravity coupling in the action.

A cross section of the deformed-NHEK horizon is still homeomorphic
to a two-sphere $S^2$, but the total area has changed.  Because the
horizon is Killing, we can compute the area along any spatial cross
section $\mathcal{H}$ carrying coordinates $x$.  The horizon
areas of the two deformed solutions are both given by
\begin{align}
  \label{eq:deformed-horizon-area}
  A_{\text{deformed}}&=\oint_{\mathcal{H}}\sqrt{\gamma}\,\mathrm{d}^2x \\ \nn
                     &=A_{\mathrm{NHEK}} \times \left[1+\eta
                       \left(\frac{\ell}{GM}\right)^4+\mathcal{O}(\varepsilon^3)\right]
                       \,,
\end{align}
where $\ell$ is $\ldcs$ or $\lgb$ when appropriate.
Here $\gamma$ is the determinant of the induced metric on
$\mathcal{H}$.  $A_\mathrm{NHEK}$ is the horizon area of an extremal
Kerr black hole, which is given by $A_\mathrm{NHEK}=8 \pi (GM)^2.$
The constant $\eta$ varies for the two deformed solutions.
For \DCS~and \DGB~respectively we find
\begin{align}
  \eta_\DCS &=  \left(4875 \sqrt{2}-1380 \pi -3928\right)/{7680} \nn \\
            &\approx -0.18 \,, \\
  \eta_\DGB & =  \left(1615 \sqrt{2}-300 \pi -464-320 \log
              2\right)/{2560} \nn\\
            &\approx +0.26 \,.
\end{align}
Despite the fact that EdGB-deformed NHEK has a true curvature
singularity, this singularity is integrable, leading to a finite
correction to the horizon area.

Note that while considering deformed NHEK, the entropy no longer
equals the horizon area, since the stringy interactions also
contribute microscopic degrees of freedom.  The horizon areas computed
here will be used in the following subsection to calculate the entropy
of the two deformed solutions.

\subsection{Thermodynamics of horizons}
\label{sec:therm-horiz}

The macroscopic entropy of a Killing horizon is interpreted as the
Noether charge associated with the Killing vector field which
generates the horizon~\cite{Wald:1993nt,Iyer:1994ys}.  In any
diffeomorphism invariant theory with a Lagrangian
$\mathscr{L}=\mathscr{L}(\phi,\nabla_a \phi,
g_{ab},R_{abcd},\nabla_eR_{abcd},\ldots)$, where $\phi$ are matter
fields, the black hole entropy can be written as an integral over a
horizon cross section $\mathcal{H}$~\cite{Jacobson:1993vj}.  Again,
since the horizon is Killing, any spacelike cross section will do.
This entropy integral is
\begin{equation}
  S = -2 \pi \oint_{\mathcal{H}}
  \frac{\delta \mathscr{L}}{\delta R_{abcd}}\hat{\epsilon}_{ab}
  \hat{\epsilon}_{cd}\bar{\epsilon}
  \,.
  \label{eq:wald-entropy}
\end{equation}
Here $\bar{\epsilon}$ is the induced volume form on the $D-2$
dimensional cross section, and $\hat{\epsilon}_{ab}$ is the
binormal.  The binormal has been normalized such that
$\hat{\epsilon}_{ab}\hat{\epsilon}^{ab}=-2$.

The NHEK solution does not have an event horizon; however, we can
still get the correct entropy of the extremal black hole by performing
the integral over the cross section of the \poinc{} horizon.
The entropy of the NHEK solution can then be obtained by evaluating
Eq.~\eqref{eq:wald-entropy} in Einstein-Hilbert theory
$\mathscr{L}=\mathscr{L}_{\mathrm{EH}}$.  It is not surprising that we
arrive at the Bekenstein-Hawking
entropy for the extremal Kerr black
hole~\cite{Bekenstein:1973ur,Hawking:1974sw},
\begin{equation}
  S_{\mathrm{NHEK}} = 2\pi\mpl^2 A_{\mathrm{NHEK}} = \frac{A_\mathrm{NHEK}}{4 G}.
\end{equation}
Similarly in \DCS{} and \DGB{}, by computing the entropy corrections
due to stringy degrees of freedom,  we will be able to obtain the 
entropies of the deformed-NHEK solutions in the two theories.
Note, however, that the entropy results agree with the extremal BH
solutions, since the \poinc{} horizon is the scaling limit of the
extremal BH event horizon.  The corrections to the entropy are due to
high-energy stringy degrees of freedom becoming activated.

In either dCS or EdGB gravity, the scalar field Lagrangian
$\mathscr{L}_{\vartheta}$ does not contribute to the entropy while the
interaction term $\mathscr{L}_{\mathrm{int}}$ does. Therefore in a
full theory with action given by Eq.~\eqref{eq:the-action}, the entropy
of a stationary black hole solution with horizon cross-section
$\mathcal{H}$ is
\begin{equation}
  S = 2\pi\mpl^2 \oint_{\mathcal{H}}\bar{\epsilon}+S_{\mathrm{int}}\,,
  \label{eq:corrected-entropy}
\end{equation}
where we have defined $S_{\text{int}}$ via
\begin{equation}
  S_{\text{int}} = -2 \pi \oint_{\mathcal{H}}
  \frac{\delta \mathscr{L}_{\text{int}}}{\delta R_{abcd}}\hat{\epsilon}_{ab}
  \hat{\epsilon}_{cd}\bar{\epsilon} \,.
\end{equation}

Compared to Einstein gravity, dCS- and EdGB-deformed NHEK receive
entropy corrections from two sources: the deformation of the horizon
area, and the string interaction term $S_{\text{int}}$.
In dCS theory, the correction to the entropy due to the scalar-gravity
interaction term is given by
\begin{equation}
  S^{\CS}_{\mathrm{int}} =
  \frac{\pi}{2}\mpl\ldcs^2 \oint_{\mathcal{H}}
  \vartheta\,{}^{*}\!R^{abcd}
  \hat{\epsilon}_{ab}\hat{\epsilon}_{cd}\bar{\epsilon}
  \,.
  \label{eq:dCS-entropy-correction}
\end{equation}
Similarly, we find the correction to entropy via the EdGB interaction is
\begin{equation}
  S^{\GB}_{\mathrm{int}} =
  \frac{\pi}{2}\mpl\lgb^2 \oint_{\mathcal{H}}
  \vartheta \, {}^{*}\!{R^*}^{abcd}
  \hat{\epsilon}_{ab}\hat{\epsilon}_{cd}\bar{\epsilon}
  \,.
  \label{eq:EdGB-entropy-correction}
\end{equation}

Now let us explore the effect of taking the decoupling limit and
compute the leading order corrections to the entropy of extremal Kerr
in \DCS\, and \DGB\, theories.  The leading order scalar field is
already at $\varepsilon^1$ while the metric perturbations correct at
order $\varepsilon^2$, thus we can evaluate
Eqs.~\eqref{eq:dCS-entropy-correction} and~\eqref{eq:EdGB-entropy-correction}
using the original NHEK metric.  Combining the horizon area
calculations given by Eq.~\eqref{eq:deformed-horizon-area}, the
entropies of the two deformed NHEK solutions can both be written as
\begin{align}
  S_{\text{deformed}} = S_{\text{NHEK}} \left[1+ \xi \left(\frac{\ell}{GM}\right)^4+\mathcal{O}(\varepsilon^3)\right]
  \,,
\end{align}
where the constant $\xi$ for \DCS~and \DGB~are given by
\begin{align}
  \xi_\DCS &= \left(4875 \sqrt{2}+360 \pi ^2-868 \pi
             -3928\right)/{7680} \nn \\
           &\approx +0.49 \,, \\
  \xi_\DGB & = \Big(360 \pi ^2+4845 \sqrt{2}-1392-960 \log 2 \nn\\
           & \quad -4 \pi  (480 \log 2-607) \Big)/7680 \nn \\
           &\approx +1.54 \,.
\end{align}
Here as well, despite the EdGB scalar solution having a singularity at
the poles, the singularity is integrable, leading to a finite
correction to the entropy.
Note that both entropy corrections are positive, as should be the case
when adding new degrees of freedom to the underlying microscopic
theory.

\section{Discussion and future work}
\label{sec:concl-future-work}

We have obtained analytic solutions for the linearized metric
deformations to near-horizon extremal Kerr spacetimes as induced by
dCS and EdGB interactions in the decoupling limit.
In this limit, the metric deformations solve linearized Einstein
equations with a source term arising from the dilaton or axion field
and the background metric.  We decomposed the metric perturbations
using basis functions adapted to the \SLTRU{} isometry, and turn the
systems of field equations into solvable ODEs.  The resulting solution
in \DCS, \emph{dCS-deformed NHEK}, is regular everywhere, while in
\DGB, \emph{EdGB-deformed NHEK} has a true curvature singularity at
the poles, discussed further below.  We
studied time-like orbits in these two newly found spacetimes.  In
particular, for circular equatorial orbits, we computed the leading
order corrections to the angular frequencies, which are observables
for sub-extremal black holes by gravitational wave experiments.
Finally, we computed the corrections to the horizon areas and the
macroscopic entropies of the extremal black hole solutions in \DCS{}
and \DGB{}.
The positive entropy corrections are related to the inclusion of new
degrees of freedom in the theory.

EdGB-deformed NHEK is irregular at the poles $u=\pm 1$, no matter how
we choose the constants of integration.  This irregular behavior
originates from the source term built from the dilaton field, since
the dilaton has an unavoidable logarithmic singularity at the poles,
as discussed in Sec.~\ref{sec:solut-scal-constr}.
This leads to a true curvature singularity, which can be seen as
follows.  We can find the singularity without solving for
$\hat{h}^{(2)}$ by simply tracing the equation of motion
Eq.~\eqref{eq:linearized-Einstein}.  Since the background Ricci scalar
and the first-order metric deformation both vanish
($\hat{R}^{(0)}=0=\hat{h}^{(1)}$), the deformation
$\delta \hat{R}^{(2)}$ is a gauge-invariant quantity.  Now, the $uu$
component of the source tensor, $S_{ab}^{\DGB}$, contains
$(\pd_{u}\hat{\vartheta})^{2}$ and $\pd_{u}^{2}\hat{\vartheta}$, which
give a pole of order two at $u=\pm 1$.  The inverse metric component
$g^{uu}$ only contributes a single zero at the poles.  Thus the trace
of the source term $g^{ab}S_{ab}^{\DGB}$ blows up with a pole of order
1 at $u=\pm1$, and we have an unavoidable curvature singularity.

This problem with extremal EdGB solutions was previously
mentioned in~\cite{Kleihaus:2011tg} and discussed further in
Appendix B of~\cite{Kleihaus:2015aje}.  They presented numerical
evidence and an analytic argument that the extremal limit does not
admit regular solutions, for any values of the GB coupling parameter.
Here, we have proven that there are no regular solutions, in the
decoupling limit.  While our analysis is restricted to the decoupling
limit, based on the gauge-invariant argument above, we have proven
that the extremal limit is indeed singular for EdGB.

We still lack a clear physical understanding of this curvature
singularity.  The simplest interpretation is that this is a sign of a
breakdown of EdGB when treated as an EFT, and that this singularity is
cured by the inclusion of operators at the same or higher order (such as those which
were discarded in the truncation of~\cite{Maeda:2009uy}).  This
situation would be a counterexample to Hadar and Reall's recent claim
that EFT does not break down at an extremal
horizon~\cite{Hadar:2017ven}.

\paragraph*{Future work.}
The near-horizon near-extremal Kerr (near-NHEK) spacetime has the same
\SLTRU{} isometry as the NHEK spacetime.  Therefore we expect all this
work can be extended to near-NHEK directly.  The techniques we used
here can also be used for any other beyond-GR theory which has a
continuous limit to GR.  Therefore, we can also solve for deformed
NHEK solutions in a broad class of theories.
It may be possible to use matched asymptotic expansions to combine
perturbation theory about (near-)NHEK and Schwarzschild, in order to
build beyond-GR metric solutions valid for all values of spin,
$0\le a \le M$.

On the observational
side, the angular frequencies of the near-extremal Kerr ISCO may be
determined accurately in future gravitational wave experiments,
providing a useful way to test general relativity.

Finally, this work
may be helpful in understanding quantum theories beyond Einstein
gravity.  We have computed the macroscopic entropies of extremal black
holes, which must be associated with corresponding microscopic
entropies.  This may be possible with an analog of the Kerr/CFT
correspondence.

\acknowledgments
The authors would like to thank
Yanbei Chen,
Jutta Kunz,
Yiqiu Ma,
Zachary Mark,
Marika Taylor,
Nicol\'as Yunes,
Peter Zimmerman,
and an anonymous referee
for useful conversations.
LCS acknowledges the support of NSF grant PHY--1404569, and
both authors acknowledge the support of the Brinson
Foundation.  Some calculations used the computer algebra
system \textsc{Mathematica}, in combination with the
\textsc{xAct/xTensor}
suite~\cite{JMM:xAct,MARTINGARCIA2008597}.

\bibliographystyle{apsrev4-1}
\bibliography{NHEK-and-stringy-corrections}

\end{document}